\documentclass[12pt,epsfig]{article}
\usepackage{epsfig}
\usepackage{amssymb}
\usepackage{amsmath}
\newcommand{\singlespace}{\renewcommand{\baselinestretch}{1}\large\normalsize}
\newcommand{\doublespace}{\renewcommand{\baselinestretch}{1.6}\large\normalsize}
\usepackage{graphicx}
\usepackage{epsfig}
\usepackage{amssymb}
\usepackage{amsmath}
\newcommand{\beq}{\begin{equation}}
\newcommand{\eeq}{\end{equation}}
\newcommand{\bea}{\begin{eqnarray}}
\newcommand{\eea}{\end{eqnarray}}

\newcommand{\ave}[1]{\langle {#1} \rangle}

\newcommand{\pslash}{p\!\!\!/}

\newcommand{\pb}{\bar\psi}
\newcommand{\qq}{\ave{\pb\psi}}
\newcommand{\eq}[1]{Eq.~(\ref{#1})}

\newcommand{\ct}{s_{22}}
\newcommand{\cf}{s_{55}}
\newcommand{\cs}{s_{77}}

\def\roughly#1{\mathrel{\raise.3ex\hbox{$#1$\kern-.75em%
\lower1ex\hbox{$\sim$}}}}

\def\gsim{\roughly>}

\def\={\;=\;}
\def\+{\;+\;}

\begin{document}

\begin{flushright}
November 2002
\end{flushright}
\vspace{1.0cm}
\begin{center}
\doublespace
\begin{large}
{\bf } Mixed phases of color superconducting quark matter
\end{large}
\vskip 1.0in
F. Neumann$^a$, M. Buballa$^{a,b}$, and M. Oertel$^{c}$\\
{\small{\it $^a$ Institut f\"ur Kernphysik, TU Darmstadt,
                 Schlossgartenstr. 9, 64289 Darmstadt, Germany\\
            $^b$ Gesellschaft f\"ur Schwerionenforschung (GSI),
                 Planckstr. 1, 64291 Darmstadt, Germany\\
            $^c$ IPN-Lyon, 43 Bd du 11 Novembre 1918,
                 69622 Villeurbanne C\'edex, France}}\\
\end{center}
\vspace{1cm}

\begin{abstract}
We examine electrically and color neutral quark matter in
$\beta$-equilibrium focusing on the possibility of mixed phases between
different color superconducting phases. To that end we apply the Gibbs
criterion to ensure phase equilibrium and discuss the external
conditions under which these mixed phases can occur. Neglecting surface
and Coulomb effects we find a rich structure of different mixed phases
with up to four components, including 2SC and CFL matter as well as
more ``exotic'' components, like a phase with $us$- and $ds$-pairing  but 
without $ud$-pairing. Preliminary estimates indicate, however, that
the mixed phases become unstable if surface and Coulomb effects are
included. 
\end{abstract}

\singlespace
\section{Introduction}
The possible existence of color superconducting matter in the core of
a neutron star is subject to intensive discussions since it has been
discovered~\cite{ARW98,RSSV98} that at intermediate densities the energy
gaps can be of the order of 100 MeV. These large values have been
obtained in the scalar color antitriplet channel where instanton
interactions~\cite{rapp00} as well as interactions derived from
single-gluon exchange predict the largest attraction. 
There are two different condensation patterns in this channel, depending 
on whether or not the strange quarks, which are more massive than the 
light up and down quarks, participate in forming a condensate,
\beq
    s_{AA'} 
    \= \ave{\psi^T \,C \gamma_5 \,\tau_A \,\lambda_{A'} \,\psi} \;.
\label{saa}
\end{equation}
Here both, $\tau_A$ and $\lambda_{A'}$ are the antisymmetric
generators of $SU(3)$, i.e., the antisymmetric Gell-Mann
matrices ($A, A'\; \in\; \{2,5,7\}$), acting in flavor and color 
space, respectively. 
In the two-flavor color superconducting phase (2SC) where only the
light quarks are involved in the condensation, the flavor
index in Eq.~(\ref{saa}) is restricted to $A = 2$. In this case 
it is always possible, without loss of generality, to perform a
color rotation such that the 2SC phase is described by $\ct \neq 0$ and 
$s_{AA'} = 0$ if $(A,A') \neq (2,2)$.  
The color-flavor-locked (CFL) phase~\cite{ARW99} is characterized by
nonvanishing condensates $\ct$, $ \cf$, and  $\cs$, i.e., it contains
$ud$ as well as $us$ and $ds$ pairs.
In the limiting case of equal masses and chemical potentials for
all quarks these three condensates are equal, whereas in general they
can be different from each other.

Most of the calculations so far have been performed using a common
chemical potential for all quarks (see, e.g.,~\cite{RaWi00,
Alford01}). It has been found that in that case a large region of the
phase diagram is occupied by a 2SC phase~\cite{BO02,OB02}. For the
interior of neutron stars, it is however important, to consider
electrically and color neutral\footnote{Strictly speaking, color
neutrality is not sufficient, but color singletness has to be
imposed. This does, however, not induce a large cost in
energy~\cite{ABMW01}, such that we can consider matter which is only
color neutral.} matter in $\beta$-equilibrium. As pointed out by
Alford and Rajagopal this strongly disfavors or even rules out the 2SC
phase in compact stars~\cite{AR02}.  For example consider a system of
massless up and down quarks together with electrons, but without
strange quarks.  Since the density of electrons is small (see, e.g.,
\cite{BO99}), to achieve electric neutrality the density of $d$-quarks
must be almost twice as large as the density of $u$-quarks, and hence
$\mu_d \approx 2^{1/3} \mu_u$. This means that, e.g., for
$\mu_u=400$~MeV, the Fermi momenta of $u$ and $d$ differ by about
100~MeV, making $ud$ BCS-pairing very difficult. The presence of
negatively charged strange quarks therefore is likely to favor pairing
even if the strange quark mass is considerably larger. Alford and
Rajagopal tackled that problem, performing an expansion in terms of
the strange quark mass. They found that, whenever the 2SC phase is
more favored than no pairing at all, the CFL phase is even more
favored.

Recently this has been reinvestigated by Steiner, Reddy and
Prakash~\cite{SRP02}, who analyzed electrically and color neutral quark matter 
within an NJL-type model, taking into account density dependent quark
masses self-consistently. For those conditions which are relevant for
neutron stars older than a few minutes, i.e., cold matter without
trapped neutrinos, the authors found the 2SC phase to be most favored 
within a small region of quark number chemical potentials. They claim,
however, that this region is likely to disappear, once the hadronic phase 
is properly taken into account~\cite{SRP02}.

The analyses of Refs.~\cite{AR02,SRP02} are based on a comparison of
homogeneous phases: For a given quark number chemical potential $\mu$ the
authors search for the phase which maximizes the pressure
after the constraints of electrical and color neutrality have been
imposed. The latter is achieved by introducing an additional chemical 
potential $\mu_Q$ which is related to electric charge and
two  chemical potentials $\mu_3$ and $\mu_8$ which are related to color.
The values of these new chemical potentials needed to neutralize
matter for a fixed value of $\mu$ depend on the phase. For instance the 
normal phase, where color invariance is not spontaneously broken, is
color neutral for $\mu_3 = \mu_8 = 0$, whereas a nonvanishing value of $\mu_Q$
is needed for electric neutrality. The situation is quite opposite
to the CFL phase where $\mu_8 \neq 0$ and $\mu_Q = \mu_3 = 0$ are 
needed~\cite{SRP02}. Finally, in the 2SC phase both, $\mu_8$ and $\mu_Q$, 
must be nonzero. Hence neutral matter in one phase is never in chemical
equilibrium with neutral matter in a different phase, even if their quark
number chemical potentials $\mu$ are the same. In particular the
points of equal pressure do not fulfill the Gibbs criteria for a 
phase transition stating that the pressure and {\it all} chemical
potentials must be the same in coexisting phases.

One possibility to resolve this problem is to give up the requirement
of separately neutral phases and to consider mixed phases of two components
in chemical equilibrium which are only neutral in total.  
This procedure has been pushed forward by Glendenning in the context 
of the quark-hadron phase transition in neutron stars 
where a similar problem occurs~\cite{G92}. Since Glendenning did not consider
color-superconducting phases, he only had to care about electric neutrality.
In this case a neutral mixed phase can obviously be constructed in those
regions of the phase boundary where the charge densities of the two
components have opposite signs.

In the present paper we generalize this procedure to construct 
electrically and color neutral mixed phases. This is more difficult
because for two charged and colored components a mixture which is 
electrically neutral is in general not color neutral and vice versa.
In the four-dimensional space spanned by $\mu$, $\mu_Q$, $\mu_3$
and $\mu_8$ the phase boundaries are three-dimensional hypersurfaces. 
As we will show the regions where electrically and color neutral mixed phases
are possible correspond to one-dimensional lines on these hypersurfaces. 

Our calculations are performed within the same NJL type model which was
the basis of the analysis of  Ref.~\cite{SRP02}. It includes
interactions which allow for the condensation patterns described
above as well as for quark-antiquark condensates $\qq$. The
latter, describing dynamical chiral symmetry breaking in the vacuum,
in particular generate dynamically a constituent strange quark mass which can
be of the order of the quark chemical potential~\cite{BO02}. 
We restrict ourselves to bulk quark matter in mean-field approximation. 
Of course if mixed phases are formed the domains belonging to a single
component must have finite sizes determined by the electric and 
color-electric fields and the surface tension. In fact, if the latter
is too large the mixed phase will become unstable and a single charged
interface between the two phases will be more favorite. This
scenario has been suggested for the transition from the hadronic phase 
to the CFL phase~\cite{ARRW01}. A crystalline color superconducting
phase~\cite{BR02} in the transition region would be another
possibility. 

The paper is organized as follows: In Sec.~\ref{charges} we introduce
the conserved charges in our system and the related chemical
potentials. We discuss the Gibbs condition for phase
boundaries. Thereafter, in Sec.~\ref{quarkmodel}, we present the model
which is used to describe quark matter. We continue in
Sec.~\ref{phasediagram} with an examination of the resulting phase
diagram in the space of the various chemical potentials. 
In Sec.~\ref{mixed} we construct electrically and color neutral mixed 
phases and discuss the resulting phase structure of neutral quark matter.
Finally, in Sec.~\ref{summary} we summarize and discuss some
open questions related to our investigations.

\section{Conserved charges, chemical potentials and phase transitions}
\label{charges}

We consider a system of quarks, electrons and muons with the number densities
$n_{f,c}, n_e, $ and $n_\mu$, respectively.
Here $f = u,d,s$ (``up'', ``down'', ``strange'')
refers to the flavor and $c = r,g,b$ (``red'', ``green'', ``blue'')
refers to the color. The total flavor and color densities are then
given by
\beq
n_f = \sum_c n_{f,c}~,\qquad n_c = \sum_f n_{f,c}~.
\eeq
As mentioned in the Introduction we are mainly interested in describing
the conditions present in compact stars older than a few minutes, 
when neutrinos can freely leave the system. 
In this case lepton number is not conserved and we have four independent
conserved charges, namely the total electric charge
\beq
n_Q =   \frac{2}{3} n_u - \frac{1}{3} n_d  -\frac{1}{3}n_s - n_e -n_\mu 
\eeq
and the three color charges $n_r$, $n_g$, and $n_b$ or,
equivalently, their linear combinations
\beq
n = n_r + n_g + n_b~,\qquad 
n_3 = n_r - n_g~,\qquad 
n_8 = \frac{1}{\sqrt{3}}(n_r + n_g - 2 n_b)~.
\eeq
Here $n$ corresponds to the total quark number density, while $n_3$
and $n_8$ describe color asymmetries. Note that $n/3$ also describes the
conserved baryon number.
The four conserved charges are related to four independent chemical 
potentials, $\mu$, $\mu_3$, $\mu_8$, and $\mu_Q$. 
The individual chemical potentials $\mu_{f,c}$ ``felt'' by the quarks 
of flavor $f$ and color $c$ are then given by corresponding components
of the diagonal matrix
\beq
\hat{\mu} = \mu \+ \mu_Q\; (\frac{1}{2} \tau_3 + \frac{1}{2 \sqrt{3}} \tau_8) 
\+ \mu_3\; \lambda_3 \+ \mu_8\; \lambda_8 ~.
\label{mus}
\eeq
Here, as before, $\tau_i$ and $\lambda_j$ are operators in flavor space
and color space, respectively. 
The electron and muon chemical potentials are simply $\mu_e =\mu_\mu = -\mu_Q$. 
This implies 
\beq
    \mu_{d,c} = \mu_{s,c} = \mu_{u,c} + \mu_e \quad \text{for all \it{c}}~,
\label{beta}
\eeq
which is usually referred to as $\beta$-equilibrium.

The total thermodynamic potential of our model is given by the sum
of a quark part and a leptonic part,
\beq
    \Omega(T,\{\mu_i\};\chi) \= 
    \Omega_q(T,\{\mu_i\};\chi) \+ \Omega_l(T,\mu_Q)~,
\eeq
where $\{\mu_i\}= \{\mu,\mu_3,\mu_8,\mu_Q\}$.
For the leptonic part we simply take a gas of noninteracting massless 
electrons and massive muons.
The quark part will be further specified in Sec.~\ref{quarkmodel}. 
Here we only note, that it depends on a set of condensates which we denote
by $\chi$. In general there are different solutions for $\chi$ at 
given temperature and chemical potentials.  Then the stable one is the 
solution which corresponds to the lowest value of $\Omega$.
At the boundaries which separate two different phases
the corresponding thermodynamic potentials coincide,
\beq
    \Omega(T,\{\mu_i\};\chi_1) \= 
    \Omega(T,\{\mu_i\};\chi_2)~, 
\label{phaseboundary}
\eeq
where $\chi_1\neq\chi_2$ if the phase transition is first order.

In a given phase the various densities defined above
can be obtained from the thermodynamic potential as
\beq
n_i = - \frac{\partial\Omega}{\partial\mu_i}~.
\label{densities}
\eeq
We are mostly interested in electrically and 
color neutral matter, which is characterized by
\beq
n_Q = n_3 = n_8 = 0~.
\label{ecneutral}
\eeq
For homogeneous phases the consequences of these constraints have been 
discussed in Refs.~\cite{AR02,SRP02}. The goal of the present article is to
analyze possible mixed phases which could exist along first-order phase 
boundaries, \eq{phaseboundary}.

Suppose we have a mixed phase consisting of two components, 1 and 2.
In general the densities $n_i^{(1)}$ and 
$n_i^{(2)}$ resulting from \eq{densities} will not be identical in both phases.
In particular, in general \eq{ecneutral} will not be fulfilled simultaneously 
for both components.  
However, as indicated in the Introduction it is sufficient to demand that 
the {\it average} charge and color densities of the mixed phase 
vanish~\cite{G92}. If the two components occupy
the volume fractions $x_1$ and $x_2 = 1-x_1$, respectively,
the average densities are given by
\beq
    n_i \= x_1\,n_i^{(1)} \+  (1-x_1)\,n_i^{(2)}~.
\label{xidef}
\eeq
This is zero for
\beq
    x_1 = \frac{n_i^{(2)}}{n_i^{(2)}-n_i^{(1)}}~.
\label{fraction}
\eeq
To be meaningful the solution must be in the interval $0<x_1<1$. 
This is fulfilled when the charge densities
$n_i^{(1)}$ and $n_i^{(2)}$ have opposite signs, which is an obvious
prerequisite for a charge neutral mixture.
For a single charge, e.g., $n_Q$, it is the only one.
However, in order to get simultaneous neutrality for three charges,
\eq{ecneutral}, we have to require that the result of \eq{fraction}
is the same for $i=Q$, 3, and 8. This is the case when
\beq
    n_Q^{(1)} \,:\, n_3^{(1)} \,:\, n_8^{(1)} \=
    n_Q^{(2)} \,:\, n_3^{(2)} \,:\, n_8^{(2)}
\label{mix}
\eeq

In our numerical calculations we will restrict ourselves to
$T=0$, relevant for compact stars older than a few minutes.  Then the
phase boundaries, \eq{phaseboundary}, are three-dimensional surfaces
in the four-dimensional space spanned by $\mu$, $\mu_Q$, $\mu_3$, and
$\mu_8$. Since \eq{mix} imposes two additional constraints,
electrically and color neutral mixed phases can be constructed along a
one-dimensional line. In the simplest case
this line starts at a point where the 
neutrality line of phase 1 ($n_Q^{(1)} = n_3^{(1)} = n_8^{(1)} = 0$), 
meets the phase boundary and it ends where the neutrality line of phase 2 
meets the phase boundary. Between these two points $x_1$ changes continuously
from 1 to 0. 
However, if the neutrality line meets another phase boundary before,
one has three coexisting phases,
\beq
    \Omega(T,\{\mu_i\};\chi_1) \= 
    \Omega(T,\{\mu_i\};\chi_2) \= 
    \Omega(T,\{\mu_i\};\chi_3)~. 
\label{phaseboundary3}
\eeq
In this case the neutrality condition reads
\beq
    \hat N\,\vec x \;\equiv\; \left (
    \begin{array}{c c c}
     n_Q^{(1)} & n_Q^{(2)} & n_Q^{(3)}\\
     n_3^{(1)} & n_3^{(2)} & n_3^{(3)}\\
     n_8^{(1)} & n_8^{(2)} & n_8^{(3)}
    \end{array}
    \right )  \left (
    \begin{array}{c}
     x_1 \\ x_2\\ x_3
    \end{array}
    \right ) \= 0~.
\eeq
In order to find a non-trivial solution for $\vec x$, we must have 
$\det \hat N = 0$. Together with \eq{phaseboundary3}, this again restricts 
the possible solutions to a one-dimensional subspace. Moreover, since the
fractions $x_i$ should not be negative, for each $m=Q,3,8$
the densities $n_m^{(i)}$ must not have
the same sign for all $i=1,2,3$.

Finally, there could even be a mixed phase, consisting of four components.
The corresponding phase boundary is one-dimensional and again the region of
possible neutral mixed phases is further restricted by the requirement
that the various fractions $x_i$ should not be negative.

\section{Quark model}
\label{quarkmodel}

The quark part of our model is defined by the Lagrangian
\beq
    {\cal L}_{eff} \= \pb (i \partial\hspace{-2.3mm}/ - \hat{m}) \psi
                      \+ {\cal L}_{q\bar q} \+ {\cal L}_{qq},
\label{Lagrange}
\end{equation}
where $\psi$ denotes a quark field with three flavors and three colors. 
The mass matrix $\hat m$ has the form 
$\hat m = diag(m_u, m_d, m_s)$ in flavor space.
Throughout this paper we will assume $m_u = m_d$,
whereas the strange quark mass $m_s$ will be different. 
To study the interplay between the color-superconducting
diquark condensates and the quark-antiquark condensates
we consider an NJL-type interaction with a quark-quark part
\beq
    {\cal L}_{qq} \=
    H\sum_{A = 2,5,7} \sum_{A' = 2,5,7}
    (\pb \,i\gamma_5 \tau_A \lambda_{A'} \,C\pb^T)
    (\psi^T C \,i\gamma_5 \tau_A \lambda_{A'} \, \psi) 
    \,.
\label{Lqq}
\end{equation}
and a quark-antiquark part
\beq
    {\cal L}_{q\bar q} \= G\, \sum_{a=0}^8 \Big[(\pb \tau_a\psi)^2
    \+ (\pb i\gamma_5 \tau_a \psi)^2\Big] 
    \;-\; K\,\Big[{\rm det}_f\Big(\pb(1+\gamma_5)\psi\Big) \,+\
                   {\rm det}_f\Big(\pb(1-\gamma_5)\psi\Big)\Big]\;.
\label{Lqbarq}
\end{equation}
Here $\tau_0 = \sqrt{\frac{2}{3}}\,1\hspace{-1.5mm}1_f$ is 
proportional to the unit matrix in flavor space. 
\eq{Lqbarq} corresponds to a typical 3-flavor NJL-model Lagrangian
which has often been used to study meson spectra 
\cite{TTKK90,KLVW90} or properties of quark matter at finite densities or
temperatures \cite{LKW92, Rehberg, BO99}.  
It consists of a $U(3)_L \times U(3)_R$-symmetric 4-point interaction and a 
't~Hooft-type 6-point interaction which breaks the the $U_A(1)$ symmetry.
The above interaction terms might arise from some underlying more 
microscopic theory and are understood to be used at mean-field level
in Hartree approximation. In particular we do not consider any contribution 
from the 6-point interaction to the diquark condensate.
The same model has also been employed in Ref.~\cite{SRP02} and 
for the case of one common chemical potential in Ref.~\cite{OB02}. 

In order to calculate the mean-field thermodynamic potential $\Omega_q$ 
at temperature $T$ and the various chemical potentials we linearize 
${\cal L}_{eff}$ in the presence of the three quark-antiquark condensates
$\phi_f = \langle \bar{f}f\rangle$, $f=u,d,s$, and the three diquark 
condensates $s_{AA}$, $A=2,5,7$. 
Introducing the constituent quark masses
\beq
    M_i \= m_i \,-\, 4G\phi_i \,+\, 2 K \phi_j \phi_k~, 
    \qquad \text{$(i,j,k)$ = any permutation of $(u,d,s)$}~, 
\end{equation}
and the diquark gaps
\begin{equation}
\Delta_A = -2 H \,s_{AA} ~,
\label{Delta12}
\end{equation}
and employing Nambu-Gorkov-formalism the result can be written in the 
following way:
\begin{alignat}{1}
    \Omega_q(T,\{\mu_i\};\chi) 
     \= &-T \sum_n \int \frac{d^3p}{(2\pi)^3} \;
    \frac{1}{2}\,{\rm Tr}\;\ln\, \Big(\frac{1}{T}\,S^{-1}(i\omega_n, \vec p)
    \Big)
    \nonumber\\
    &+\, 2G\,(\phi_u^2 \,+\, \phi_d^2 \,+\, \phi_s^2) 
     \;-\; 4K \phi_u\,\phi_d\,\phi_s 
    \+ H\,(|\ct|^2 \,+\, |\cf|^2 \,+\, |\cs|^2)\,. 
\label{Omega}
\end{alignat}
Here $\chi = \{\phi_u,\phi_d,\phi_s,\ct,\cf,\cs\}$ denotes the set of 
condensates.
$S^{-1}$ is the inverse fermion propagator 
\beq
    S^{-1}(p) \= \left(\begin{array}{cc}
    \pslash - \hat{M} + \hat{\mu}\gamma^0 &
    \Delta_2 \gamma_5\tau_2\lambda_2 + 
    \Delta_5 \gamma_5\tau_5\lambda_5 + 
    \Delta_7 \gamma_5\tau_7\lambda_7  \\
    -\Delta_2^* \gamma_5\tau_2\lambda_2
    -\Delta_5^* \gamma_5\tau_5\lambda_5 - 
     \Delta_7^* \gamma_5\tau_7\lambda_7  &
    \pslash - \hat{M} - \hat{\mu}\gamma^0
    \end{array}\right)\,,
\label{Sinv}
\end{equation}
where $\hat{M} = diag(M_u,M_d,M_s)$. In \eq{Omega}, it has to be 
evaluated at $p=(i\omega_n, \vec p)$, where $\omega_n = (2n-1)\pi T$ are 
fermionic Matsubara frequencies. 

For the case of equal chemical potentials an explicit expression for
$\Omega_q$ has been discussed in detail in Ref.~\cite{BO02}. 
For unequal chemical potentials this is more involved
and we only state the general structure as far as necessary for our 
later discussion. 
As shown in Refs.~\cite{AR02,SRP02,ABR99} $S^{-1}$ can be decomposed 
into several blocks. Six of them have a $2\times 2$ structure in the
combined color-flavor space,
\beq
{\cal M}_{f_1 c_1, f_2 c_2} = 
\left(\begin{array}{cc} \pslash^+_{f_1,c_1} -M_{f_1} &
\Delta \gamma_5 \\ - \Delta^* \gamma_5 & \pslash^-_{f_2,c_2} -
M_{f_2}\end{array}\right)~,
\label{block2}
\eeq
where $\pslash^\pm_{f,c} = \pslash \pm \mu_{f,c} \gamma_0$.
These blocks describe pairing of two species of quarks with flavors
$f_1$ and $f_2$ and colors $c_1$ and $c_2$, respectively. 
More precisely, one gets the following three combinations \cite{SRP02}:
First, green up quarks are paired with red down quarks, where 
$\Delta=\Delta_2$, second, blue down quarks are paired with green strange 
quarks ($\Delta=\Delta_5$), and third, red strange quarks are paired with 
blue up quarks ($\Delta=\Delta_7$). 
The other three $2\times 2$ blocks can be generated simply by interchanging 
the role of particles and antiparticles and lead to the same combinations.

The remaining contribution to $\Omega$ has the following 
$6\times 6$-structure:
\beq
{\cal M}_{uds} = \left(
\begin{array}{cccccc}\pslash^+_{u,r}-M_u&0&0&0&-\Delta_2\gamma_5
&-\Delta_5\gamma_5 \\ 
0&\pslash^+_{d,g}-M_d&0&-\Delta_2\gamma_5&0&
-\Delta_7\gamma_5 \\ 
0&0&\pslash^+_{s,b}-M_s&-\Delta_5\gamma_5&
-\Delta_7\gamma_5&0 \\
0&\Delta_2^*\gamma_5&\Delta_5^*\gamma_5&\pslash^-_{u,r}-M_u&0&0\\ 
\Delta_2^*\gamma_5&0&\Delta_7^*\gamma_5&0&\pslash^-_{d,g}-M_d&0\\ 
\Delta_5^*\gamma_5&\Delta_7^*\gamma_5&0&0&0&\pslash^-_{s,b}-M_s
\end{array} \right)~,
\label{block6}
\eeq
involving red $u$-quarks, green $d$-quarks and blue $s$-quarks.
This represents the particle and antiparticle contributions to the
$3\times 3$-block mentioned in Ref.~\cite{ABR99}.  

As demonstrated by Rajagopal and Wilczek for a simplified two-quark model,
the pairing of two quark species forces their densities to be 
equal~\cite{RW01}. Originally, it was expected that in the CFL phase, 
where all quarks are paired, all densities $n_{f,c}$ are equal. This 
would mean that the CFL phase is always electrically and color neutral.
However, as pointed out by Steiner, Reddy, and Prakash~\cite{SRP02}, 
only those quarks which are paired in the same  $2\times 2$ block have the 
same density, whereas the densities could differ for different blocks.
Furthermore the argument does not apply to the $6\times 6$ block.
For the CFL phase this means
$n_{u,g} = n_{d,r}$, $n_{u,b} = n_{s,r}$, and $n_{d,b} = n_{s,g}$,
and thus 
\beq
n_{u} = n_{r}~, \qquad n_{d} = n_{g}~,\qquad  
n_{s} = n_{b}\qquad \text{(CFL)}~.
\label{denscfl}
\eeq
This relation guarantees neutrality of CFL matter under the rotated
electromagnetism $\tilde Q$~\cite{AR02,SRP02}, but in general it does
not preclude the presence of ordinary electric or color charges.
Note, however, that color neutral CFL matter is automatically electrically 
neutral as long as no leptons are present.

In the 2SC phase, where $\Delta_5 = \Delta_7 = 0$, \eq{block6} can be
further decomposed, and we obtain a new $2\times 2$ block involving
red $u$-quarks and green $d$-quarks. Together with the other  
$2\times 2$ block which contains $\Delta_2$ this leads to the relations
\beq 
n_{u,r} = n_{d,g}~,\qquad n_{u,g} = n_{d,r} \qquad \text{(2SC)}~.
\label{dens2sc}
\eeq 
The corresponding relations for other phases, e.g., with two nonvanishing
diquark condensates, can be obtained analogously.

The further elaboration of the thermodynamic potential contains only
straight forward manipulations, which will not be presented here. 
In \eq{Omega} $\Omega_q$ depends on $\chi$, i.e., on the choice of the
condensates  $\phi_f$ and $s_{AA}$. In order to find the self-consistent
solutions we have to look for the stationary points of the potential with 
respect to these condensates. As mentioned before, there is often more
than one selfconsistent solution. In this case the stable solution is the
one which minimizes the thermodynamic potential.

Our model is incomplete without specifying the parameters to be used in 
the numerical calculations. 
To this end we adopt the parameters of Ref.~\cite{Rehberg}, which have
been determined by fitting the pseudoscalar meson masses and decay constants
in vacuum: a sharp 3-momentum cutoff $\Lambda = 602.3$~MeV
to regularize the integrals, the bare quark masses $m_u = m_d = 5.5$~MeV and 
$m_s = 140.7$~MeV, and the coupling constants $G = 1.835 \Lambda^{-2}$
and $K = 12.36 \Lambda^{-5}$.
Similarly, one could fix the quark-quark coupling constant $H$ by fitting
baryon masses within a Fadeev approach \cite{IBY93,HaKr95}.
For simplicity, however, we follow
Ref.~\cite{SRP02} (preprint versions v1 and v2)
and take $H = G$.\footnote{In the latest version of Ref.~\cite{SRP02} (v3),
the authors take $H = 3/4\,G$. Qualitatively, this does not change their
results. Therefore, in the following we will not distinguish between the
different versions. However, for any quantitative comparison with our
results, the reader should refer to version v2.}  
This is similar in magnitude 
to the value we have chosen in Ref.~\cite{BO02} where we took
$H = 3/4\,G$, but with a larger $G$\footnote{In 
Ref.~\cite{BO02}
we used a four-point interaction with the quantum numbers of a single-gluon 
exchange, without six-point interaction. In this case a simple Fierz 
rearrangement yields $H:G = 3:4$. This relation also holds for 
a two-flavor instanton interaction. Hence, in our present model
a simple estimate for the value of the diquark interaction, e.g., in the 
$ud$-channel, is given by $H = 3/4\,G_{eff}$, where 
$G_{eff} = G - K\ave{\bar s s}$. 
In vacuum this is slightly larger than $G$, but it is a density dependent
quantity and differs for different phases.
Such effects have not been taken into account in our present work, but
they are certainly worthwhile to be studied in future.}.

\section{Phase structure}
\label{phasediagram}

In order to construct the equation of state for electrically and color 
neutral quark matter (including possible mixed phases) we first have to
explore the phase structure of the model and the charge densities 
as functions of the chemical potentials $\mu_i$.

\begin{figure}
\begin{center}
\epsfig{file=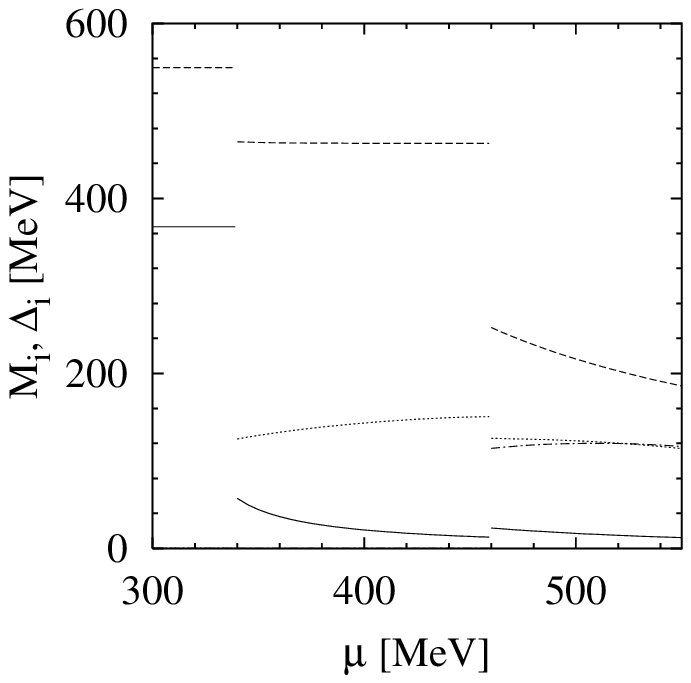,width=7.cm}
\epsfig{file=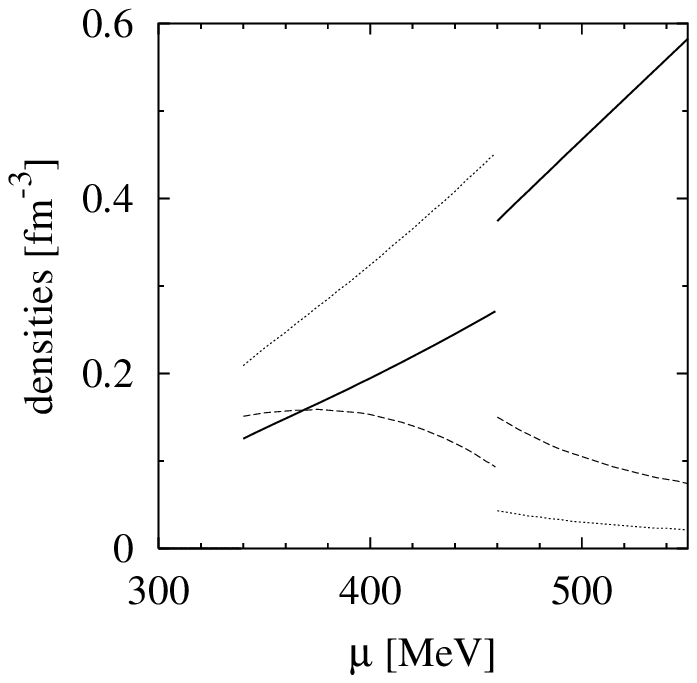,width=7.cm}
\end{center}
\vspace{-0.5cm}
\caption{\small Gap parameters and densities at $T =\mu_3 = \mu_8 = \mu_Q = 0$
as functions of $\mu$. Left: $M_u=M_d$ (solid), $M_s$ (dashed), $\Delta_2$
(dotted), and $\Delta_5=\Delta_7$ (dashed-dotted). Right: $n/10$ (solid),
$n_8$ (dashed), and $n_Q$ (dotted).
}
\label{fig0}
\end{figure}

We begin with the standard case of a single, color and flavor independent, 
chemical potential $\mu$ for all quarks, i.e., $\mu_3 = \mu_8 = \mu_Q = 0$. 
For the parameters specified above we obtain the
results which are displayed in Fig.~\ref{fig0}. In the left panel the
constituent masses and diquark gaps are shown. One can clearly distinguish
three phases separated by first-order phase boundaries: 
At low chemical potentials we find a normal phase with vanishing diquark 
condensates, followed by the 2SC phase with $\Delta_2 \neq 0$ 
and finally the CFL phase where also $\Delta_5$ and $\Delta_7$ are nonzero. 
At the phase boundaries we observe strong discontinuities in the quark 
masses~\cite{BO02,OB02}.

In the right panel we show the corresponding densities. Note that the
quark number density $n$ (solid line) has been divided by 10 to fit to the
scale. 
The dotted line corresponds to the electric charge density $n_Q$, the 
dashed line to the color density $n_8$. The color density $n_3$ is identically
zero.
As one can see, all densities vanish in the ``normal phase'', i.e., this
phase corresponds to the vacuum. In fact, it has to be like this, because
as soon as up and down quarks are present, their Fermi surfaces are
subject to a Cooper instability leading to the formation of the diquark
condensate $\ct$. (This argument will no longer go through, once we have 
switched on one of the other chemical potentials which lift the degeneracy 
of the Fermi surfaces of all up and down quarks.)

The two other phases carry both, electric and color charges. The electric
charge of the 2SC phase is easily understood. Since $\mu_Q = 0$, there are
no leptons and the densities of up and down quarks are equal. Moreover,
in our example
there are no strange quarks, which are too heavy to be populated in this
regime. Hence the total electric charge density is given by $n_Q = n/6$. 
The nonvanishing color density $n_8$ reflects the fact that for equal
chemical potentials the densities of the paired (red and green) quarks are 
larger than the density of the unpaired (blue) quarks~\cite{BHO02}.
Numerically we find $(n_r-n_b)/n = 10\%$ at the lower boundary and
$(n_r-n_b)/n = 3\%$ at the upper boundary of the 2SC phase.

Just above the transition to the CFL phase this ratio does not change
very much, whereas the electric charge density drops significantly due to a
strong increase of the density of strange quarks.
To a large extent, this is caused by a sudden drop of the strange quark mass,
but this is only part of the story. 
For instance, at $\mu = 500$~MeV we have $M_u = M_d = 17.2$~MeV and $M_s = 216.5$~MeV.
Using these numbers in a free gas approximation we would expect
$n_Q = 0.049$~fm$^{-3}$, whereas numerically we find $n_Q = 0.030$~fm$^{-3}$.
This difference is caused by the diquark pairing, which links the 
flavor densities in the CFL phase directly to the color densities, as 
discussed in \eq{denscfl}. For $n_3 = 0$ one finds 
$n_Q = 1/(2\sqrt{3})\,n_8$, in agreement with our numerical results.

Electrically and color neutral CFL and 2SC matter has been constructed
in Ref.~\cite{SRP02}. According to the charge densities $n_8$ and $n_Q$ 
which have to be removed, this in general requires the introduction
of nonvanishing chemical potentials $\mu_8$ and $\mu_Q$.
In order to see how these additional chemical potentials influence
the phase boundaries, we first study them separately. 

\begin{figure}
\begin{center}
\epsfig{file=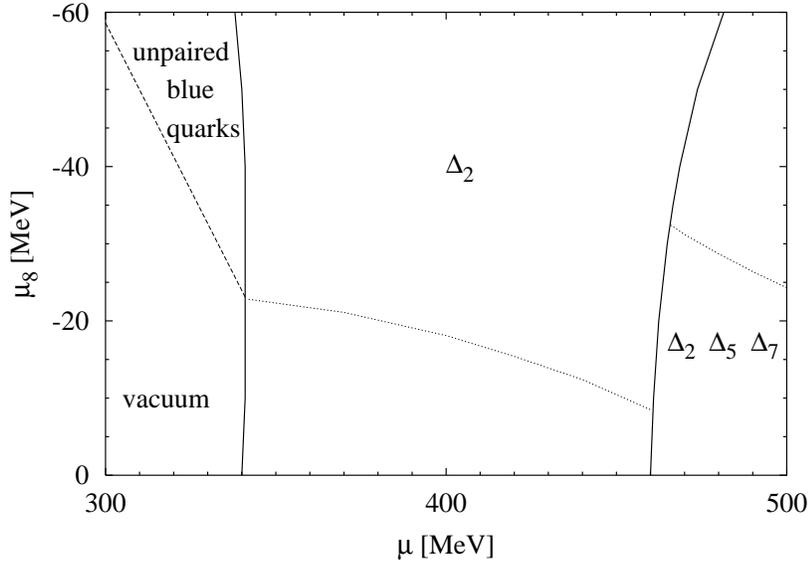,width=11.cm}
\end{center}
\vspace{-0.5cm}
\caption{\small Phase diagram in the $\mu-\mu_8$ plane for 
$T = \mu_3 = \mu_Q = 0$. 
The various phases separated by the solid lines are  characterized by 
different non-vanishing diquark gaps $\Delta_i$ as indicated in the figure.
In the non-superconducting phase quarks are present only above the dashed 
line. The dotted lines are the lines of color neutral matter. In the 
CFL phase this also corresponds to electrically neutral matter.
}
\label{phasemu8}
\end{figure}

In Fig.~\ref{phasemu8} we show the phase diagram in 
the $\mu-\mu_8$-plane for $\mu_Q = \mu_3 = 0$. The (first-order) phase
boundaries are indicated by solid lines. We find again the three
phases discussed before, i.e., the normal phase, the 2SC phase, and the CFL 
phase. For $\mu_8 = 0$ we have seen that the ``normal phase'' actually 
corresponds to the vacuum. However, when $\mu_b = \mu - 2/\sqrt{3} \mu_8$
becomes larger than the vacuum masses of the light quarks (the region above 
the dashed line), blue up and down quark states can be populated forming 
a gas of unpaired blue quarks. Here we have neglected that in principle
these quarks could pair in a different channel~\cite{Sch00,BHO02b}.
In any case we should keep in mind that our model is not suited for a
realistic description of the low-density regime, where confinement and 
hadronic degrees of freedom have to be taken into account.

For the color superconducting phases we have also indicated the lines of 
color neutral matter (dotted). As discussed in the previous section, 
in the CFL phase color neutral quark matter is automatically electrically 
neutral as well, i.e., in the CFL phase the dotted line corresponds to the 
neutral matter solution, which has been determined in Ref.~\cite{SRP02}.
It meets the phase boundary to the 2SC phase at
$\mu = 465.7$~MeV and $\mu_8 = -32.5$~MeV. The 2SC matter which is in 
chemical and mechanical equilibrium with the neutral CFL matter at this
point carries both, electric and color charge, $n_Q = 0.464$~fm$^{-3}$ and
$n_8 = -0.329$~fm$^{-3}$. In the next section, this point 
will be the end point of the 2SC-CFL mixed phase.
Unlike color neutral CFL matter, color neutral 2SC matter is not
electrically neutral but positively charged. In fact, a nonvanishing 
$\mu_8$ does not change the ratio of up and down quarks and hence,
as long as no strange quarks are present, $n_Q/n = 1/6$ as before.

\begin{figure}
\begin{center}
\epsfig{file=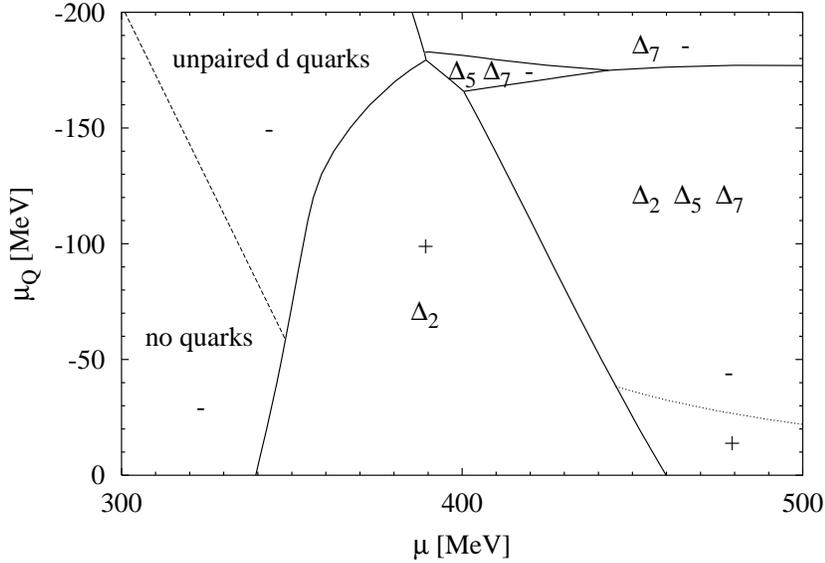,width=11.cm}
\end{center}
\vspace{-0.5cm}
\caption{\small Phase diagram in the $\mu-\mu_Q$ plane for 
$T = \mu_3 = \mu_8 = 0$. 
The various phases separated by the solid lines are  characterized by 
different non-vanishing diquark gaps $\Delta_i$ as indicated in the figure.
In the non-superconducting phase quarks are present only above the dashed 
line.
The ``+'' and ``-'' signs indicate the sign of the electric charge
density in the corresponding region. 
The dotted line corresponds to electrically (but not color)
neutral matter in the CFL phase. 
}
\label{phasemuq}
\end{figure}

In Fig.~\ref{phasemuq} we show the phase diagram in 
the $\mu-\mu_Q$-plane for $\mu_8 = \mu_3 = 0$. 
Since we are interested in neutralizing the electrically positive
2SC phase, we choose $\mu_Q$ to be negative.
As long as this is not too large, we find again the normal phase at lower
values of $\mu$, the 2SC phase in the intermediate region and the CFL phase
for large $\mu$. This changes dramatically around $\mu_Q \simeq 180$~MeV
where both, the 2SC phase and the CFL phase disappear and a new phase 
emerges. This phase is analogous to the 2SC phase but with $ds$ pairing, 
instead of $ud$ pairing (``2SC$_{ds}$''). In a small intermediate regime
there is even another phase which contains $us$ and $ds$ but no $ud$
pairs (``SC$_{us+ds}$''). 

Qualitatively, the existence of these phases is quite plausible: 
At low values of $|\mu_Q|$ the Fermi momenta of the up and down quarks 
are relatively similar to each other, whereas the strange quarks are suppressed
because of their larger mass. With increasing negative $\mu_Q$, however,
the up quarks become more and more disfavored and eventually the 
Fermi momenta are ordered as $p_F^u < p_F^s < p_F^d$. It is then 
easy to imagine that only $ds$ pairing or -- in some intermediate regime --  
$us$ and $ds$ pairing is possible. 

\begin{figure}
\begin{center}
\epsfig{file=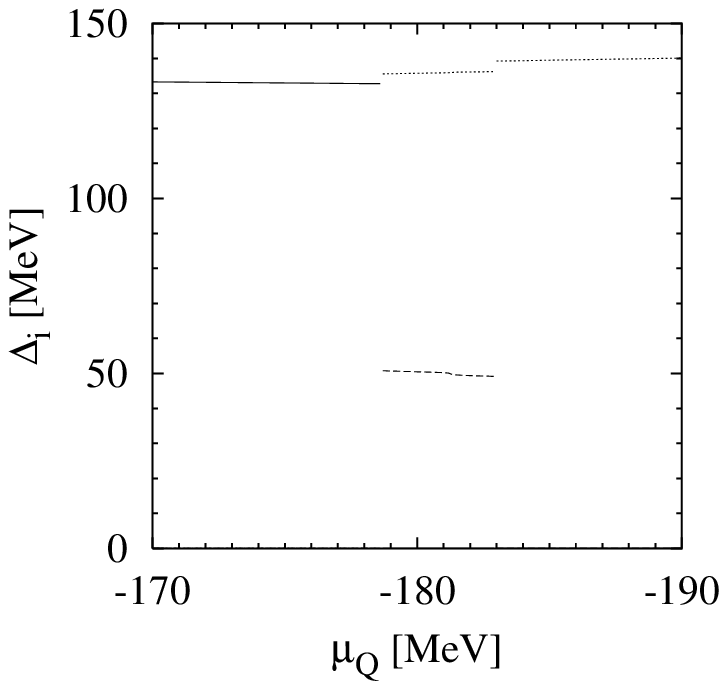,width=7.cm}
\epsfig{file=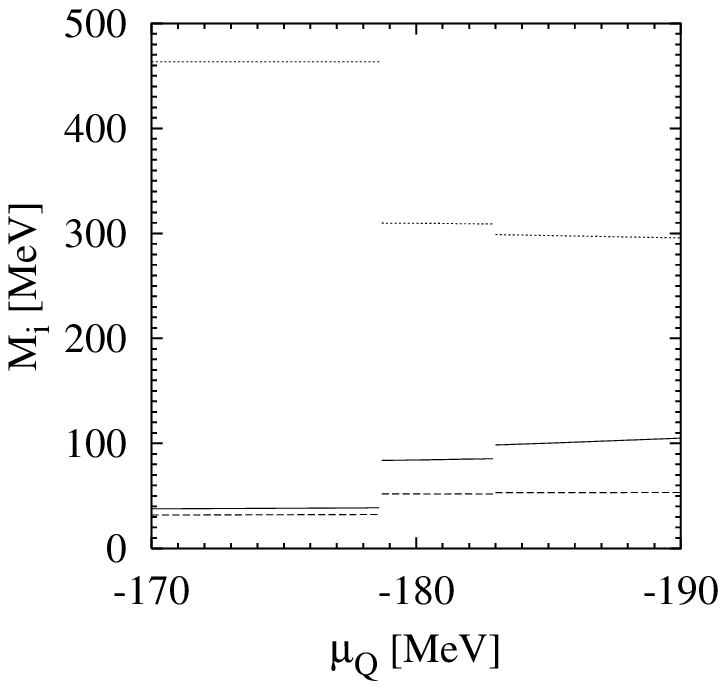,width=7.cm}
\end{center}
\vspace{-0.5cm}
\caption{\small Diquark gaps and quark masses for
$T =\mu_3 = \mu_8 = 0$, and $\mu = 390$~MeV as functions of $\mu_Q$. 
Left: $\Delta_2$ (solid), $\Delta_5$ (dashed), and $\Delta_7$ (dotted).
Right: $M_u$ (solid), $M_d$ (dashed), and $M_s$ (dotted).
}
\label{gapsq}
\end{figure}

Following this argument one might expect that there is always a value
of $\mu_Q$, where the Fermi momenta of up and strange quarks are equal 
and hence the 2SC phase should be either followed by the CFL phase or by
a phase with $us$-pairing only. However, this is not the case
because of the discontinuous behavior of the quark masses.
This is illustrated in Fig.~\ref{gapsq} where the
diquark gaps and constituent quark masses are shown as functions of
$\mu_Q$ for fixed $\mu = 390$~MeV and  $\mu_3 = \mu_8 = 0$.
The 2SC--SC$_{us+ds}$ phase transition takes place at $\mu_Q = -178.6$~MeV,
corresponding to $\mu_u = \mu + 2/3 \mu_Q \simeq 270$~MeV and
$\mu_d = \mu_s = \mu - 1/3 \mu_Q \simeq 450$~MeV. 
Below the transition point the strange quark mass is even larger than 
460~MeV and, consequently, no strange quarks are present.
At the transition point the strange quark mass drops to 310~MeV and 
the nominal Fermi momentum $p_F^s = \sqrt{\mu_s^2-M_s^2}$
is immediately larger than $p_F^u$.

The stability of the various condensates is rather well described by
the criterion~\cite{RW01}
\beq
    \Delta_{ij} \;\gsim\;  \left|\frac{p_F^i - p_F^j}{\sqrt{2}}\right| \;=:\;
    \Delta_{ij}^c~,
\label{stab}
\eeq
where $\Delta_{ij}$ is the gap related to the pairing of the quark species
$i$ and $j$. In the 2SC phase just below the phase boundary we have
$\Delta_2 = 132.8$~MeV, slightly larger than $\Delta_2^c =127.4$~MeV. 
At the phase boundary $\Delta_2^c$ rises to 133.6~MeV due to a sudden
increase of the up quark mass by more than 40~MeV. Taking the earlier
value of $\Delta_2$, the above criterion is no longer fulfilled, which
is consistent with our numerical result that the $ud$-pairs break up.
This level of agreement is certainly better than one should
expect. In fact, in the SC$_{us+ds}$ phase we find $\Delta_5$ continuously
decreasing from 50.8~MeV to 49.1~MeV whereas $\Delta_5^c$ increases
from 48.2~MeV to 52.6~MeV, slightly violating \eq{stab}. It is nevertheless
possible to understand the break-up of the $us$ pairs, which occurs at
$\mu_Q = -183.0$~MeV, from the fact that at this point $\Delta_5^c$ jumps 
to 62.6~MeV due to a further increase of $M_u$ and a further decrease
of $M_s$. Moreover, the fact that we always find
$\Delta_5 \approx \Delta_5^c$, at least in this example, indicates that
the SC$_{us+ds}$ phase is rather fragile and might disappear upon small
variations of the model parameters.

In the phase diagram, Fig.~\ref{phasemuq}, we also indicated the
sign of the electric charge density for the various regions,
and the line of electrically neutral matter in the CFL phase (dotted line). 
Note that there is no other electrically neutral regime in this diagram
(apart from the vacuum at small $\mu$ and $\mu_Q = 0$). In the normal phase, 
there are again no quarks below the dashed line, corresponding to the line
$\mu - 1/3 \mu_Q = M_d$. This region is nevertheless negatively charged
due to the leptons which are present for any $\mu_Q < 0$. Above the dashed 
line there are also $d$ quarks rendering the matter even more negative.
(In the right corner of this phase there is also a very small fraction
of $u$ quarks.)

\begin{figure}
\begin{center}
\epsfig{file=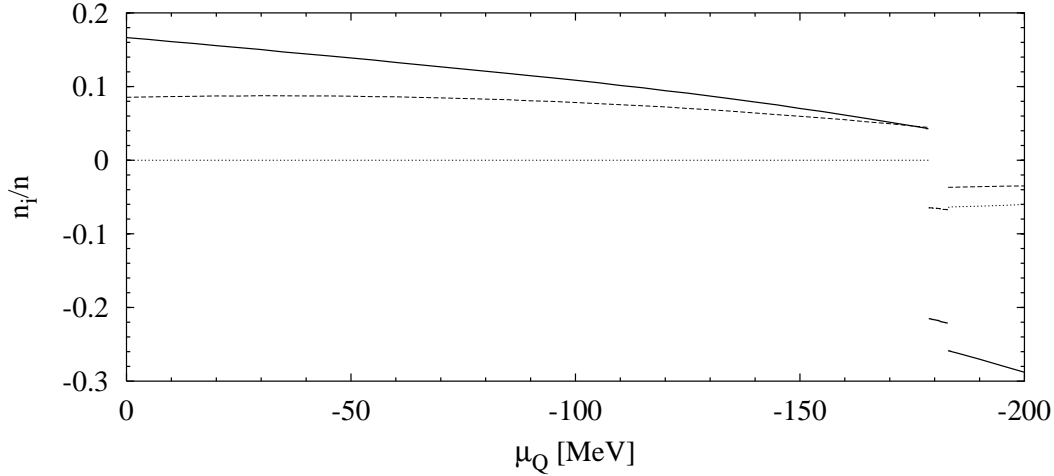,width=14.cm}
\end{center}
\vspace{-0.5cm}
\caption{\small Relative densities for
$T =\mu_3 = \mu_8 = 0$, and $\mu = 390$~MeV as functions of $\mu_Q$: 
$n_Q/n$ (solid), $n_8/n$ (dashed), and $n_3/n$ (dotted). 
}
\label{densq}
\end{figure}

The ``new'' phases, 2SC$_{ds}$ and SC$_{us+ds}$, are also negatively charged.
On the contrary, the entire 2SC phase is positively charged, even at the
largest values of $|\mu_Q|$. This is illustrated in Fig.~\ref{densq}
where the various charge densities $n_i$ divided by the total quark
number density $n$ are plotted as functions of $\mu_Q$, again for 
fixed $\mu = 390$~MeV and $\mu_3 = \mu_8 = 0$. As expected, $n_Q/n$ 
(solid line) decreases with increasing negative $\mu_Q$. However, in
the 2SC phase ($0 \ge \mu_Q > -178.6$~MeV) it stays positive and
before the point of neutrality is reached the phase transition to the 
SC$_{us+ds}$ phase takes place. 

\begin{table}[b]
\begin{center}
\begin{tabular}{|c|c|c|c|c|c|c|c|c|}
\hline
phase        &  N  & 2SC & 2SC$_{us}$ & 2SC$_{ds}$  & SC$_{ud+us}$
             & SC$_{ud+ds}$ & SC$_{us+ds}$ & CFL  \\ \hline
diquark gaps & --- & $\Delta_2$ & $\Delta_5$ & $\Delta_7$ 
             & $\Delta_2$, $\Delta_5$ & $\Delta_2$, $\Delta_7$ 
             & $\Delta_5$, $\Delta_7$ 
             & $\Delta_2$, $\Delta_5$, $\Delta_7$
\\ \hline
\end{tabular}
\end{center}
\caption{\small Phases and corresponding non-vanishing diquark gaps. 
}
\label{table1}
\end{table}  
The difficulty to obtain electrically neutral 2SC matter can be traced
back to the fact that according to \eq{dens2sc}, the sum of red and
green $u$ quarks is equal to the sum of red and green $d$ quarks. As
long as no strange quarks are present, the related positive net charge
can only be compensated by the blue quarks and the leptons, which
would require a very large negative $\mu_Q$. However, before this
point is reached it becomes more favorite to form a different phase
with a relatively large fraction of strange quarks which then also
participate in a diquark condensate.\footnote{This is very similar to
the arguments of Alford and Rajagopal~\cite{AR02}. The main difference
is that we do not compare different {\it neutral} phases with each
other, but phases in chemical equilibrium.}  As discussed above, the
selfconsistent treatment, which leads to a sudden drop of the strange
quark mass and hence to a sudden increase of the strange Fermi
momentum, is crucial in this context.

In the above discussion we have not considered the effect of a non-vanishing
chemical potential $\mu_8$ on top of a non-vanishing $\mu_Q$.
Since the blue quarks are the main carriers of negative 
electric charge in the 2SC phase, one could hope that increasing the 
number of blue quarks, as necessary for color neutrality, 
could also help to electrically neutralize 2SC matter.  
It turns out, however, that the rather small values of $\mu_8$ which are
needed for color neutrality (see Ref.~\cite{SRP02} and
Fig.~\ref{phasemu8})
do not change the above results qualitatively. 

So far we always kept $\mu_3 = 0$. In fact, there is no need to
vary $\mu_3$, as long as we are mainly interested in finding 
electrically and color neutral solutions of homogeneous normal, 2SC, or CFL 
matter. However, as we will see in the next section, the construction
of neutral mixed phases requires also non-vanishing 
values of $\mu_3$. In this context we will encounter another phase,
which is not present in Figs.~\ref{phasemu8} and \ref{phasemuq}.
For illustration we consider a plane in the four-dimensional 
$\{\mu_i\}$-space where $\mu$ and $\mu_Q$ are taken as independent 
variables and $\mu_3$ and $\mu_8$ are given by
$\mu_3 = -\mu_Q/2$ and $\mu_8 = -\mu_Q/7 -30$~MeV.
The relevance of this particular choice will become more clear in the next
section. Here we just note that $\mu_3 = -\mu_Q/2$ means that
$\mu_{u,r} = \mu_{d,g}$ .
Also the sum $\mu_{s,r}+\mu_{u,b}$, corresponding to the chemical potential
related to a pair of a red strange quark and a blue up quark, equals the
sum $\mu_{s,g}+\mu_{d,b}$, corresponding to the chemical potential
related to a pair of green strange quarks and blue down quarks.
Together with the relations given above \eq{denscfl} and the isospin
symmetry of the original Lagrangian this implies for the CFL phase
that $n_u = n_d$ or, equivalently, $n_r = n_g$ and thus $n_3 = 0$.

\begin{figure}
\begin{center}
\epsfig{file=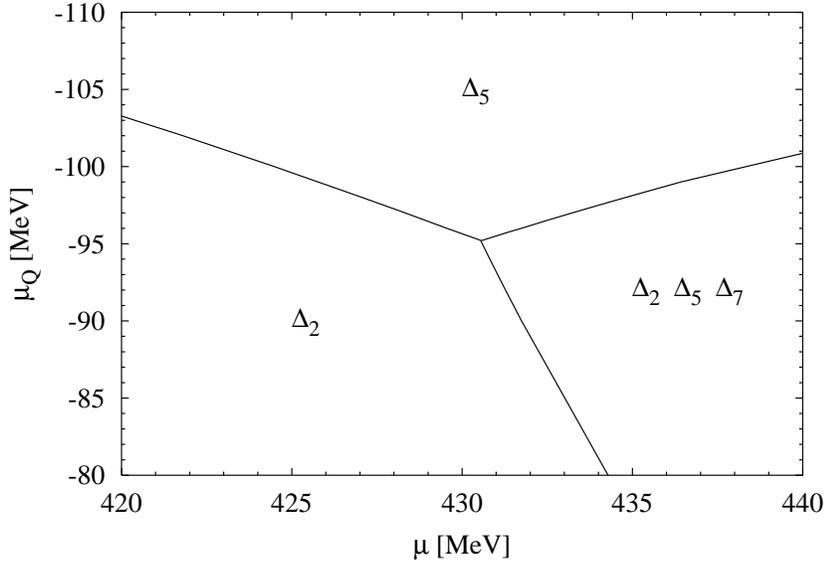,width=11.cm}
\end{center}
\vspace{-0.5cm}
\caption{\small Phase diagram in a plane defined by $\mu$ and $\mu_Q$ as 
independent variables and $\mu_3 = -\mu_Q/2$ and $\mu_8 = -\mu_Q/7 -30$~MeV.
The various phases separated by the solid lines are  characterized by 
different non-vanishing diquark gaps $\Delta_i$ as indicated in the figure.
}
\label{phase3}
\end{figure}

In Fig.~\ref{phase3} we show a small part of the resulting phase diagram.
Here, in addition to the standard 2SC and CFL phases, we find a phase
where only $u$ and $s$ quarks are paired (``2SC$_{us}$'').
This means, we have found examples for all three  
$\Delta_A$, $A=2,5,7$, being the only non-vanishing scalar diquark gaps
in some regime. Taking all possible combinations of no, one, two, or three
of these condensates (see Table~\ref{table1}),
the phases SC$_{ud+us}$ and SC$_{ud+ds}$, i.e., 
the combinations $\Delta_2+\Delta_5$ and $\Delta_2+\Delta_7$ are 
the only ones we have not encountered. These phases might exist as well,
but we did not perform a systematic investigation of this question,
which was not the goal of this work.

\section{Mixed phases}
\label{mixed}

\begin{figure}
\begin{center}
\epsfig{file=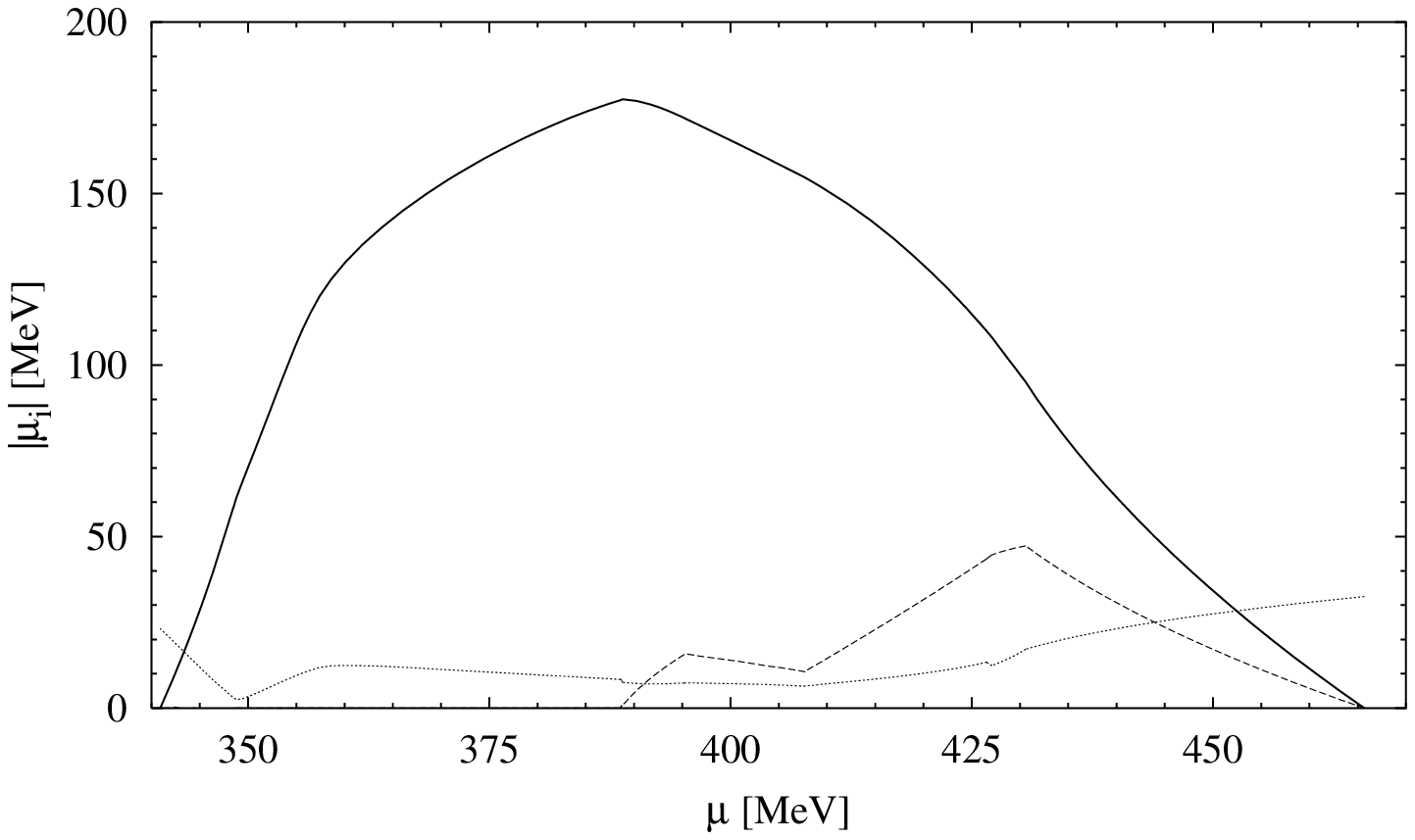,width=14.cm}
\end{center}
\vspace{-0.5cm}
\caption{\small Chemical potentials $\mu_i$ corresponding to the 
                electrically and color neutral mixed phases, listed in
                Table~\ref{table2}: 
                $-\mu_Q$ (solid), $\mu_3$ (dashed), and $-\mu_8$ (dotted).
}
\label{mixmutot}
\end{figure}

Having gained some overview about the general phase structure of our
system we are now in the position to construct electrically and color
neutral mixed phases.  Our starting point is $\mu = 465.7$~MeV, $\mu_8
= -32.5$~MeV, and $\mu_3 = \mu_Q = 0$ where the line of neutral CFL
matter meets the boundary to the 2SC phase. At lower values of $\mu$,
mixed phases become possible and are energetically favored as long as
Coulomb and surface effects are neglected. Following the guidelines
worked out in Sec.~\ref{charges} we find nine regimes characterized by
different compositions of coexisting phases (see Table~\ref{table2}).
The corresponding chemical potentials $\mu_i$ as functions of $\mu$
are displayed in Fig.~\ref{mixmutot}.
In Table~\ref{table2} we also list the corresponding minimal and 
maximal quark number densities, averaged over the components of the
respective mixed phase.

\begin{table}[b!]
\begin{center}
\begin{tabular}{|l|c|c|}
\hline
components & $\mu$ [MeV] & $n$ [fm$^{-3}$] \\ 
\hline
N, 2SC & 340.9 - 388.6  &0.00 - 1.56\\ \hline
N, 2SC, SC$_{us+ds}$& 388.6 - 388.7  & 1.56 - 1.56\\ \hline
N, 2SC, SC$_{us+ds}$, 2SC$_{us}$& 388.7 - 388.8 & 1.56 - 1.56\\ \hline
2SC, SC$_{us+ds}$, 2SC$_{us}$& 388.8 - 395.4 & 1.56 - 1.74 \\ \hline
2SC, SC$_{us+ds}$& 395.4 - 407.7 & 1.74 - 2.59\\ \hline
2SC, SC$_{us+ds}$, CFL & 407.7 - 426.5 & 2.59 - 2.90\\ \hline
2SC, SC$_{us+ds}$, CFL, 2SC$_{us}$& 426.5 - 427.1  & 2.90 - 2.93  \\ \hline
2SC, CFL, 2SC$_{us}$& 427.1 - 430.6  &2.93 - 3.11  \\ \hline
2SC, CFL& 430.6 - 465.7  &3.11  - 3.92
\\ \hline
\end{tabular}
\end{center}
\caption{\small Composition of electrically and color neutral mixed phases,
                corresponding quark number chemical potentials and average
                quark number densities. The various components are defined 
                in Table~\ref{table1}.}
\label{table2}
\end{table}  

\begin{figure}
\begin{center}
\epsfig{file=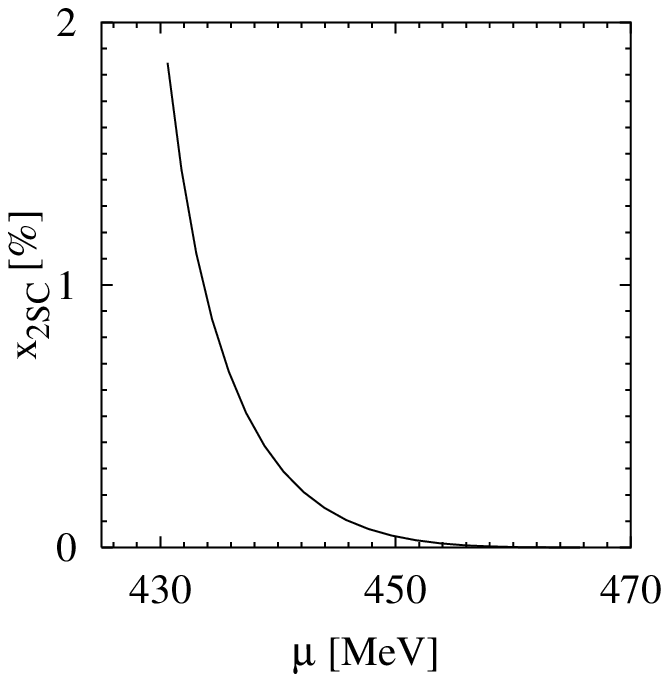,width=7.cm}\qquad
\epsfig{file=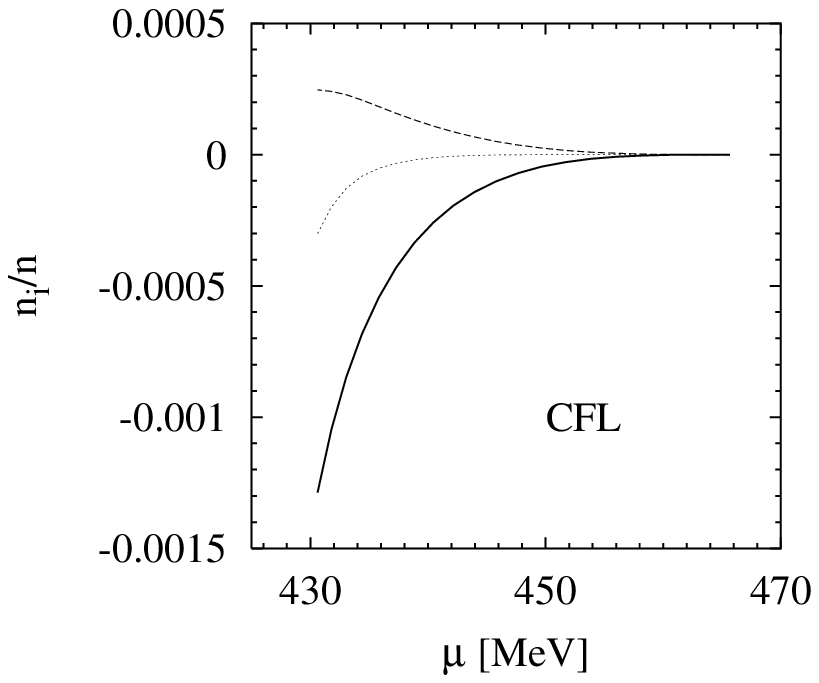,width=7.cm}
\end{center}
\vspace{-0.5cm}
\caption{\small Quantities related to the neutral 2SC-CFL mixed phase as 
functions of the quark number chemical potential $\mu$. 
Left: Volume fraction $x_{2SC}$ of the 2SC component.
Right: Relative densities in the CFL component:
$n_Q/n$ (solid), $n_8/n$ (dashed), and $n_3/n$ (dotted). 
}
\label{mixmux}
\end{figure}

In the regime closest to the region of homogeneous neutral CFL matter 
(430.6~MeV~$< \mu <$~465.7~MeV), we find a mixed phase consisting of
a CFL component and a 2SC component. The volume fraction $x_{2SC}$
of the 2SC component is displayed in the  left panel of Fig.~\ref{mixmux}.
In the higher-$\mu$ part of this region it is completely negligible, but
even at the lower end it remains below 2\%. 
Consequently, the CFL component must stay 
almost neutral by itself. Indeed, the relative charge densities 
$n_i/n$, $i=3,8,Q$, (right panel of Fig.~\ref{mixmux}) are very small.
As we have discussed in the previous section, $n_3$-neutrality of the 
CFL phase 
is maintained by the relation $\mu_3 = -\mu_Q/2$. For the actual values of
$\mu_3$ and $\mu_Q/2$ in the 2SC-CFL mixed phase we find a deviation 
of less than 1\% from this relation, while $\mu_8$ can approximately be
fitted by $\mu_8= -\mu_Q/7 - 30$~MeV. This is the reason why we have
calculated the phase diagram
shown in Fig.~\ref{phase3} with these constraints.

As can be seen
there, the 2SC-CFL phase boundary meets the boundary to
the 2SC$_{us}$ phase at $\mu = 430.6$~MeV and we get a three-component
neutral mixed phase, consisting of 2SC, CFL and 2SC$_{us}$.  Below
that, on a short interval in $\mu$, we even find a four-component
neutral mixed phase (2SC, CFL, 2SC$_{us}$, and SC$_{us+ds}$) before
upon further decreasing $\mu$ the system goes over into a neutral
2SC-CFL-SC$_{us+ds}$ mixed phase.

\begin{figure}
\begin{center}
\epsfig{file=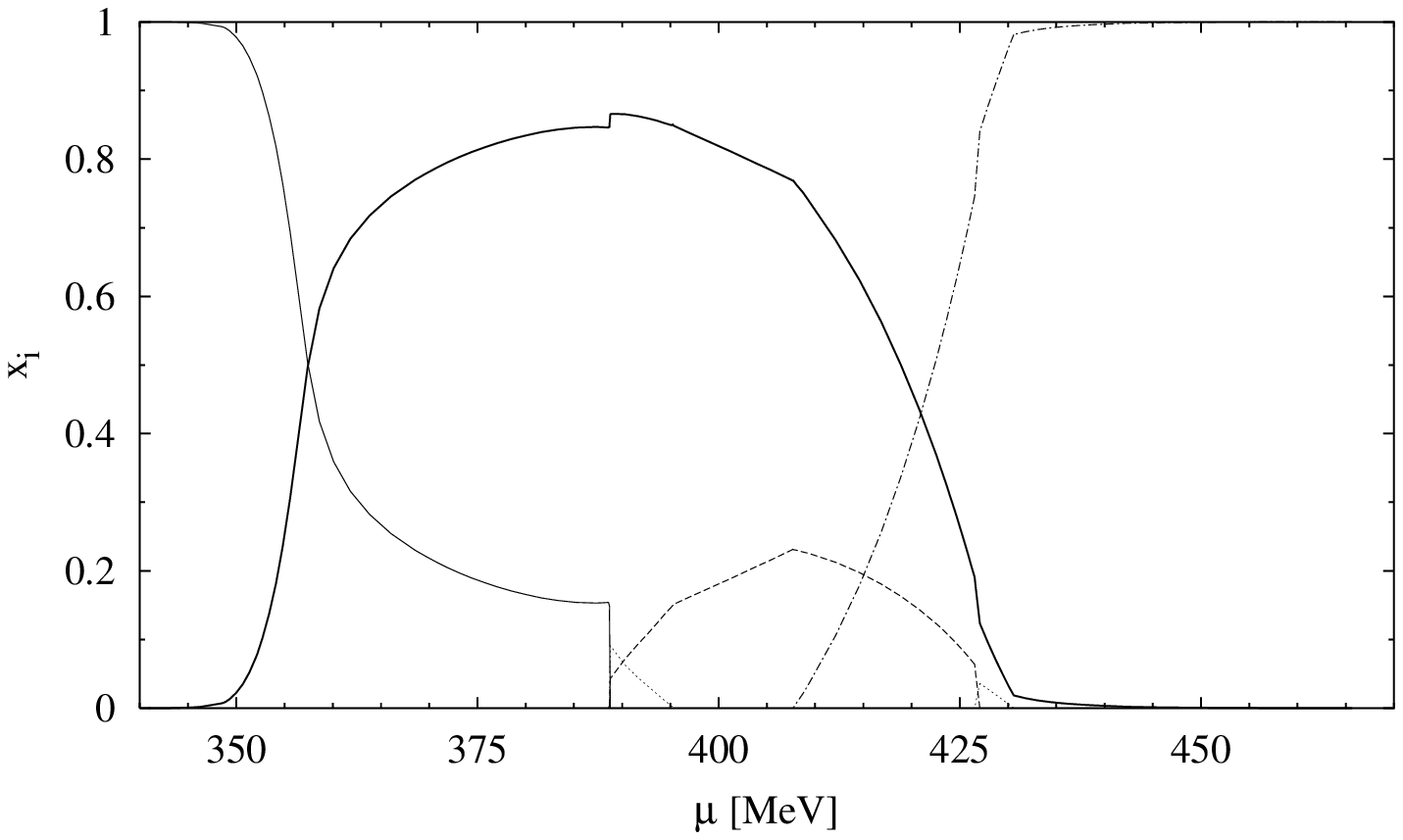,width=14.cm}
\end{center}
\vspace{-0.5cm}
\caption{\small Volume fractions of the various components in the 
         mixed phase region as functions of $\mu$: normal (thin solid),
         2SC (thick solid), CFL (dashed-dotted), SC$_{us+ds}$ (dashed),
         2SC$_{us}$ (dotted).
}
\label{mixxtot}
\end{figure}

In Fig.~\ref{mixxtot} the volume
fractions of the various components of the mixed phases are plotted as
functions of $\mu$.
Whereas the 2SC-CFL-mixed phase (Fig.~\ref{mixmux}) is completely
dominated by the CFL component, thereafter the CFL fraction becomes
quickly smaller with decreasing $\mu$, while in particular the 2SC
component, becomes more and more important.
At $\mu = 407.7$~MeV the CFL component disappears completely.

An admixture of normal quark matter is found below $\mu =
388.8$~MeV. The fractions of the superconducting phases other than the
2SC phase then rapidly become smaller and vanish at $\mu = 388.6$~MeV,
while the fraction of normal matter strongly increases.\footnote{Note,
however, that all volume fractions $x_i$ are continous functions in the
entire mixed phase region.} 
Nevertheless the 2SC phase stays the dominant component for
$\mu\gsim 360$~MeV. 

As discussed in the previous section, apart from the vacuum there is 
no region of stable neutral non-superconducting quark matter. Therefore the
normal-2SC mixed phase cannot end in normal homogeneous quark matter 
but only in the vacuum. To that end, the chemical potential $\mu_Q$ must 
finally go to zero. Eventually, at $\mu = 348.6$~MeV, $|\mu_Q|$ drops below
60~MeV and we enter the regime where the normal phase only consists of
electrons without quarks (cf. Fig.~\ref{phasemuq}). This means the 
corresponding mixed phase consists of (electrically positive) 2SC-droplets 
surrounded by regions without quarks and neutralized by a homogeneous 
background of electrons. Since the electrons are color neutral, the
2SC component must be color neutral by itself. This is maintained
by an increase of $|\mu_8|$ in this regime. 
At $\mu = 340.9$~MeV we finally reach the vacuum.

Qualitatively our results are in agreement with those obtained in
Ref.~\cite{SRP02}, where no mixed phase was admitted. The regions
which are dominated by the CFL phase and the 2SC phase, respectively,
correspond roughly to those values of the chemical potential where the
authors of Ref.~\cite{SRP02} find that the CFL or the 2SC phase,
respectively, is favored.  
Of course, since we started from the same Lagrangian, these homogeneous 
neutral solutions also exist in our model. We find, for
instance, below $\mu = 465.7$~MeV a color and electrically neutral CFL
phase which is, however, less favored than the 2SC-CFL mixed phase.
At this point we should recall that in our calculations we have
neglected the surface energy and the energies of the electric and 
color-electric fields. Clearly the mixed phases are only stable if the 
gain in bulk free energy is larger than these so-far neglected contributions.
If this is not the case we recover the results of Ref.~\cite{SRP02}.
In a neutron star we would then find homogeneous phases separated by sharp 
interfaces.

For the two-component mixed phases the surface and Coulomb contributions
can be estimated along the lines of the analysis performed in 
Ref.~\cite{ARRW01} for the interface between nuclear and CFL matter.
Because of the smallness of the 2SC fraction in the 2SC-CFL mixed phase 
the gain in bulk free energy compared with a pure CFL phase is extremely 
small in this case (less than 0.1~MeV/fm$^3$). It is therefore more or less 
obvious that this mixed phase will not be stable.
The situation looks a little more promising for the 2SC-SC$_{us+ds}$-phase 
at $\mu \approx 400$~MeV where the rarer phase, i.e., the SC$_{us+ds}$-phase,
is more populated. Here we find a gain in bulk energy of about 4~MeV/fm$^3$.
However, even in this case a calculation along the lines of 
Ref.~\cite{ARRW01} shows that this gain is weight out by 
Coulomb and surface energy already for relatively small values of the surface 
tension $\sigma\approx 10$~MeV/fm$^2$. Note that the true value of $\sigma$ 
is an unknown quantity which does not follow from our model. However, as 
argued in Ref.~\cite{ARRW01} it is likely to be much larger than the
above value.

For mixed phases with three or four components the situation is obviously
more complicated. It seems,
however, rather unlikely that an admixture of more than two components
can be favored if the surface and Coloumb energy already inhibits the
existence of a phase with two components. The more components we have
the more contributions to surface, Coulomb and color charge energy we
have. On the other hand, even if, e.g., a 3-component mixed phase turns out
to be unstable, we cannot exclude to find a stable 2-component mixed phase 
instead. A definite answer to these questions lies beyond the scope of this 
paper and we postpone it to future work.

\section{Summary and Discussion}
\label{summary}
We performed an investigation of quark matter under conditions which
possibly occur in the interior of neutron stars older than a few
minutes, i.e., dense, electrically and color neutral matter in
$\beta$-equilibrium.  Recently it was shown by Alford and Rajagopal
that these constraints have important effects, in particular that they
might disfavor the standard two-flavor color superconducting phase
(2SC) in neutron stars~\cite{AR02}. In the present article we focussed
on the formation of mixed phases, i.e., different coexisting phases in
chemical equilibrium.  In this case the neutrality conditions have
only to be fulfilled globally~\cite{G92}.

Our model calculations are performed within an NJL-type model, which
allows to treat diquark condensates on an equal footing with
quark-antiquark condensates, leading to density (and phase-) dependent
effective quark masses~\cite{BO02}.  In accordance with our
expectations the regimes of lower or higher quark number chemical
potentials are dominated by normal (non-superconducting) matter and by
matter in the color-flavor locked (CFL) phase, respectively.  In
addition, we find a certain regime of quark number chemical potentials
where the 2SC phase is the dominant component.  Naively one would
expect that the above three phases represent the only constituents of
the mixed phases we encounter.  However, in addition we find
admixtures of more exotic phases, like, e.g., a SC$_{us+ds}$-phase or
a 2SC$_{us}$-phase. These phases occur because the chemical
potentials, in particular $\mu_Q$, necessary to enforce electric
neutrality, are often large enough such that the ordering of the Fermi
momenta is reversed: $p_F^u<p_F^s<p_F^d$.

Simple estimates show that the gain in bulk energy due to the formation
of these mixed phases is probably not large enough to compensate for the 
energy arising from surface, and electric and color charge effects. 
This seems to indicate that homogeneous phases of neutral quark matter 
separated by sharp interfaces are favored. This could include the
possibility of a stable neutral 2SC phase. However, a more careful analysis
is certainly needed before a definite answer to this question can be given. 
There are other aspects which deserve to be studied more closely in future:

At low densities we certainly need a more realistic description of the
hadronic phase, which in the present model at best is represented by
a vacuum or quark phase with spontaneously broken chiral symmetry.
A more complete calculation, including a realistic equation of state of
nuclear matter, has to show whether in particular 
the N-2SC-phase which we find at rather low densities is
robust. 

At higher densities we should also take into account the
possibility of a CFL-phase with kaon condensation~\cite{BS02,KR02}. 
Recently the authors of Ref.~\cite{SRP02} have estimated the 
effect of this possibility on the equation of state of homogeneous
neutral quark matter. They found that the kaon condensate lowers the 
critical quark chemical potential for the 2SC-CFL phase transition
by about 16~MeV. This is not dramatic but could make a difference
if for instance the transition point to the hadronic phase turns out 
to be in the same region.   

We also neglected the possibility of the formation
of crystalline (LOFF) phases~\cite{BR02,ABR01}. These become
eventually favored over phases with BCS pairing if the Fermi surfaces
are pulled apart. As we are mostly dealing with pairing of quarks with
unequal Fermi momenta, we suppose that especially in the vicinity of
phase boundaries, the possible formation of a crystalline phase should
be taken into account. 
In fact, the occurrence of mixed phases which are necessarily related to
non-uniform structures in space already points into this direction. 
The neglect of gradient terms in our mean-field approach is 
only justified if the structures are much larger than the average distance
between the quarks. 
Unfortunately, we cannot say much about the sizes in the mixed phases as long
as we do not know the surface tension. However, taking relatively small
surface tensions as required to keep the mixed phases energetically stable,
the sizes of the rarer components are typically only a few times larger 
than the inter-quark distances.

Recently another interesting pairing configuration has been discussed
by Liu and Wilczek~\cite{LW02}.
A so-called ``interior gap'' can be formed in
systems with unequal Fermi momenta where the Fermi momentum of the
heavier fermion species is larger than that of the lighter
species. The authors of Ref.~\cite{LW02} claim that the ``interior
gap'' configuration is favored
if
\beq
1> \frac{p_F^a+p_F^b}{2 p_F^a} \frac{M_a}{M_a+M_b}~,
\label{interior}
\eeq
where $M_a<M_b$ and $p_F^a<p_F^b$. 
Usually this possibility can be excluded in studies of
superconducting quark matter, since the Fermi momentum of the heavier
strange quarks is smaller than the Fermi momenta of up and down
quarks. But, as we have discussed above, the
existence of ``exotic'' phases, like, e.g., the SC$_{us+ds}$-phase,
can be traced back to the fact that the Fermi momenta are ordered as
$p_F^u < p_F^s < p_F^d$. Therefore, $us$-pairing in the ``interior
gap'' configuration seems possible. Let us look at some numerical
example: At $\mu \approx 400$~MeV we find SC$_{us+ds}$ matter with
$M_u\approx 60$~MeV, $M_s \approx 300$~MeV, $p_F^u\approx 285$~MeV,
and $p_F^s\approx 350$~MeV. This leads to values of about
0.2 for the
right hand side of \eq{interior}. Thus it seems probable that
such a phase occurs.

It has also to be mentioned that we have restricted ourselves to 
diquark condensates $s_{AA'}$ with $A=A'$. Originally this was 
motivated by the fact that, e.g., in the 2SC phase any linear combination
of $s_{22}$, $s_{25}$, and $s_{27}$ can be rotated into the 22-direction
by a global color transformation. However, in principle we
can always construct a color neutral mixed phase without applying
color chemical potentials by combining several
components of the same condensate but rotated into different color 
directions. Within our present formalism this would further lower 
the free energy, but it is rather unlikely, whether this is still
the case when the color electric energy of the various domains are
taken into account. Anyway, problems, like the difficulty to
find stable 2SC matter, mostly arise from the constraint of electric
neutrality, whereas imposing color neutrality is only a minor effect.

Note that we only considered the situation relevant for neutron stars
older than a few minutes, i.e., when the neutrinos can freely leave the
star. The authors of Ref.~\cite{SRP02} performed an extended analysis
looking also at a system at finite temperature and lepton content.
These complications are likely to be relevant during the evolution of
the star from a proto-neutron star at high temperature where neutrinos
are trapped to a cold compact star where neutrinos can freely leave
the star. Steiner, Reddy and Prakash~\cite{SRP02} showed that in this
case the 2SC phase is favored because the CFL phase excludes
electrons and can therefore not easily accommodate a finite lepton
number chemical potential. A study of mixed phases in this
context is left to future work.

\section*{Acknowledgments}
Two of us (M.B. and M.O.) would like to thank K. Rajagopal for
illuminating discussions and the ITP, Santa Barbara, for their
hospitality and financial support (Grant
No. PHY94-07194). M.O. acknowledges financial support from the
Alexander von Humboldt-foundation as a Feodor-Lynen fellow.


\begin{thebibliography}{99}
\itemsep=0cm
\bibitem{ARW98}  M. Alford, K. Rajagopal, and F. Wilczek, 
                 Phys. Lett. B 422 (1998), 247.
\bibitem{RSSV98} R. Rapp, T. Sch\"afer, E.V. Shuryak, and M. Velkovsky,
                 Phys. Rev. Lett. 81 (1998) 53.
\bibitem{rapp00} R. Rapp, T. Sch\"afer, E.V. Shuryak, and
                 M. Velkovsky, Annals Phys. 280 (2000) 35.
\bibitem{ARW99}  M. Alford, K. Rajagopal, and F. Wilczek,
                 Nucl. Phys. B 537 (1999) 443.
\bibitem{RaWi00} K. Rajagopal and F. Wilczek, hep-ph/0011333,
                 and references therein.
\bibitem{Alford01} M. Alford, Ann. Rev. Nucl. Part. Sci. 51 (2001) 131.
\bibitem{BO02}   M. Buballa and M. Oertel, Nucl. Phys. A 703 (2002) 770.
\bibitem{OB02}   M. Oertel  and M. Buballa,  
                 in: {\it Ultrarelativistic heavy ion collisions},
                 Proc. Intl. workshop XXX on gross properties of nuclei 
                 and nuclear excitations, Hirschegg/Austria,
                 M. Buballa, W. N\"orenberg, B.-J. Schaefer, 
                 and J. Wambach (eds.), Darmstadt (2002).
\bibitem{ABMW01} P. Amore, M.C. Birse, J.A. McGovern, and N. R. Walet,
Phys. Rev. D 65 (2002) 074005.
\bibitem{AR02}   M. Alford and K. Rajagopal, JHEP 0206 (2002) 031.
\bibitem{BO99}   M. Buballa and M. Oertel, Phys. Lett. B 457 (1999) 261.
\bibitem{SRP02}  A. Steiner, S. Reddy, and M. Prakash, hep-ph/0205201.
\bibitem{G92}    N.K. Glendenning, Phys. Rev. D 46 (1992) 1274.
\bibitem{ARRW01} M. Alford, K. Rajagopal, S. Reddy, and F. Wilczek,
                 Phys. Rev. D 64 (2001) 074017.
\bibitem{BR02}   J. Bowers and  K. Rajagopal, Phys. Rev. D 66 (2002), 065002.
\bibitem{TTKK90} M. Takizawa, K. Tsushima, Y. Kohyama and K. Kubodera,  
                 Nucl. Phys. A 507 (1990) 611.
\bibitem{KLVW90} S. Klimt, M. Lutz, U. Vogl and W. Weise, 
                 Nucl. Phys. A 516 (1990) 429; \\
                 U. Vogl, M. Lutz, S. Klimt and W. Weise, 
                 Nucl. Phys. A 516 (1990) 469.
\bibitem{LKW92}  M. Lutz, S. Klimt and W. Weise, 
                 Nucl. Phys. A 542 (1992) 521.
\bibitem{Rehberg} P. Rehberg, S.P. Klevansky and J. H\"ufner,
                  Phys. Rev. C 53 (1996) 410.
\bibitem{ABR99}  M. Alford, J. Berges, K. Rajagopal, Nucl. Phys. B 558 (1999)
                 219.
\bibitem{IBY93}  N. Ishii, W. Bentz, and K. Yazaki, 
                 Phys. Lett. B 318 (1993) 26.
\bibitem{HaKr95} C. Hanhart and S. Krewald, Phys. Lett. B 344 (1995) 55.
\bibitem{BHO02}  M. Buballa, J. Ho\v sek, and M. Oertel,
                 Phys. Rev. D 65 (2002) 014018.
\bibitem{RW01} K. Rajagopal and F. Wilczek, Phys. Rev. Lett. 86 (2001), 3492.
\bibitem{Sch00}  T. Sch\"afer, Phys. Rev. D 62 (2000)  094007.
\bibitem{BHO02b} M. Buballa, J. Ho\v sek, and M. Oertel, hep-ph/0204275.
\bibitem{ABR01} M. Alford, J. Bowers, and K. Rajagopal, Phys. Rev. D
63 (2001) 074016.
\bibitem{LW02} W. V. Liu and F. Wilczek, cond-mat/0208052.
\bibitem{BS02} P.F. Bedaque and T. Sch\"afer, Nucl. Phys. A 697 (2002) 802.
\bibitem{KR02} D.B. Kaplan and S.Reddy, Phys. Rev. D 65 (2002) 054042.
\end{thebibliography}
\end{document}